\documentstyle[amssymb,aps,12pt]{revtex}
\begin{document}
\title{{\bf Charged Vacuum Bubble Stability}}
\author{J.R. Morris}
\address{{\it Physics Dept., Indiana University Northwest,}\\
{\it 3400 Broadway, Gary, Indiana 46408}\\
\bigskip\ }
\maketitle

\begin{abstract}
A type of scenario is considered where electrically charged vacuum bubbles,
formed from degenerate or nearly degenerate vacuua separated by a thin
domain wall, are cosmologically produced due to the breaking of a discrete
symmetry, with the bubble charge arising from fermions residing within the
domain wall. Stability issues associated with wall tension, fermion gas, and
Coulombic effects for such configurations are examined. The stability of a
bubble depends upon parameters such as the symmetry breaking scale and the
fermion coupling. A dominance of either the Fermi gas or the Coulomb
contribution may be realized under certain conditions, depending upon
parameter values.

\smallskip\ 

\end{abstract}


\section{Introduction}

Domain wall formation\cite{vilenkin,vsbook,ktbook} can result from the
spontaneous breaking of a discrete symmetry, such as a $Z_2$ symmetry. If,
however, the discrete symmetry is {\it biased}\cite{clo}, where the
formation of one protodomain is favored over that of the other, (or if the
discrete symmetry is {\it approximate}\cite{fggk,ggk}, rather than exact, so
that the probabilities of forming domains of different vacua become
unequal), then there can result a network of bounded domain wall surfaces.
This network of ``{\it vacuum bubbles}'' will evolve in a way that is
dictated by the interactions between the scalar field giving rise to the
domain wall and other fields. A vacuum bubble formed from an ordinary domain
wall not coupled to other particles or fields undergoes an unchecked
collapse due to the tension in the wall. However, if there is a coupling
between the domain wall scalar field and one or more fermions, it is
possible for the fermions to help stabilize the bubble. For instance, an
interaction term $G_F\phi \bar{\psi}\psi $, where $\phi $ represents the
scalar field forming the domain wall, generates a mass $m_F$ for the fermion 
$\psi $. If $\phi \rightarrow \pm \eta $, $m_F\rightarrow G_F\eta $
asymptotically outside the wall and $\phi \rightarrow 0$ in the core of the
wall, then it becomes energetically favorable for the fermions to populate
the core of the wall, with the fermions experiencing an attractive force $%
\vec{F}\sim -G_F\nabla \phi $. A thin walled vacuum bubble may then feel a
force, due to the existence of an effective two-dimensional Fermi gas
pressure, that tends to slow or halt the collapse of the bubble. This type
of effect plays an essential role in the ``Fermi ball'' model\cite{mca95},
for example, where heavy neutral fermions acquire mass from the domain wall
scalar field. The resulting bag configuration in the Fermi ball model can
possess a finite equilibrium surface area, but, because the bag is unstable
against flattening, the bag can flatten and subsequently fragment into many
tiny, thick walled Fermi balls that are supported by the Fermi gas. Fermi
balls can serve as candidates for cold dark matter, and such a model can
arise quite naturally in supersymmetric theories in response to softly
broken supersymmetry\cite{mb}.

Here, a similar type of model is considered where the fermions populating
the bubble wall have an electric U(1) gauge charge, and attention is focused
upon the effects of the domain wall tension, the Fermi gas, and the
Coulombic forces on the stability of an electrically charged vacuum bubble.
In this type of model, the long range Coulombic force tends to stabilize a
thin walled bubble against flattening and fragmenting, so that the final
equilibrium configuration can consist of a larger, thin walled, charged
vacuum bubble instead of the smaller Fermi ball. For a bubble populated by
only one species of fermion, the relative importance of the Coulombic and
the Fermi gas contributions depends upon the number of fermions in the
bubble. Furthermore, as pointed out in sec. 2, the stability of the charged
bubble against fermion emission and charge evaporation is found to depend on
the strengths of model parameters such as the fermion coupling constant $G_F$
and the symmetry breaking scale $\eta $. A bubble that is not stable against
charge evaporation, due to the existence of a critical strength electric
field, can cause electron-positron pairs to be produced near its surface.
Either electrons or positrons are then attracted to the bubble surface to
partially or completely neutralize it, with the result that the vacuum
bubble ends up supporting two species of fermions, which increases the Fermi
gas contribution while decreasing the Coulombic one. Such vacuum bubbles
inhabited by two fermion species are examined in sec. 3. Specifically, we
look at completely neutralized bubbles and near-critically charged bubbles
(with near-critical surface electric fields). Bubble sizes and masses are
estimated for limiting cases of special interest, and parameter constraints
are estimated. A brief summary of the results is presented in sec. 4.

\section{Charged Vacuum Bubbles with a Single Fermion Species}

As stated in the introduction, if the mass of an electrically charged
fermion is generated by a scalar field, as might be described by the
interaction term $G_F\phi \bar{\psi}\psi $, and the scalar field forms
domain walls where $\phi =0$ in the core of a wall, then it becomes
energetically favorable for the fermion to reside inside the wall where it
is massless. A domain wall then acquires electrical charge due to a
population of charged fermions. Consider the case where the domain wall
arises in response to a broken $Z_2$ symmetry, with the domain wall
interpolating between two distinct, but energetically degenerate, vacuum
states. If the discrete $Z_2$ symmetry is biased\cite{clo}, so that the
probabilities of forming domains of different vacua become unequal, then
there can result a network of bounded domain wall surfaces which may evolve
to give rise to stable or metastable charged vacuum bubbles. (We could also
consider the case where the $Z_2$ symmetry is approximate, with the
difference in vacuum energy densities being sufficiently small that it can
be safely ignored.) Let us focus on a single spherical domain wall bubble
that encloses vacuum (say, $\phi =\mp \eta $) and is surrounded by vacuum
(say, $\phi =\pm \eta $). In the thin wall approximation, the bubble can be
considered as a two dimensional surface populated by massless, electrically
charged fermions. The bubble is then considered to carry a uniform charge $%
q=Ne$ (but no spatial electric current). The configuration energy ${\cal E}$
of the bubble receives contributions from the surface energy $\Sigma $ of
the wall, the two dimensional Fermi gas energy ${\cal E}_F$ , and the
Coulomb self energy ${\cal E}_C$. The new feature in this model is the
inclusion of a long ranged U(1) gauge field which can help to stabilize the
bubble against collapse and fragmentation, as takes place in the
electrically neutral Fermi ball model, for instance. A model of the type
under consideration here might be described by a Lagrangian of the form 
\begin{equation}
L=\frac 12(\partial \phi )^2+\bar{\psi}(i\gamma \cdot D-G_F\phi )\psi -\frac 
14(F^{\mu \nu })^2-\frac{\lambda ^2}2(\phi ^2-\eta ^2)^2.  \label{e1}
\end{equation}

A planar domain wall solution is given by $\phi (x)=\eta \tanh (x/w)$, where 
$w=1/(\lambda \eta )$ is the thickness, or width, of the wall. The wall has
a surface energy of $\Sigma =\frac 43\lambda \eta ^3$. We consider a
spherical bubble of domain wall inhabited by $N\gg 1$ fermions, each of
charge $e$, which, in the thin wall approximation, has a radius $R\gg w$,
i.e. $\lambda \eta R\gg 1$. (For simplicity, we shall normally take $\lambda
\sim 1$.) The configuration energy of the bubble is 
\begin{equation}
{\cal E=E}_W+{\cal E}_F+{\cal E}_C,  \label{e2}
\end{equation}

\noindent where ${\cal E}_W=\Sigma S$ is the energy contribution from the
domain wall, with $S=4\pi R^2$ the surface area, 
\begin{equation}
{\cal E}_F=\frac{4\sqrt{\pi }N^{3/2}}{3\sqrt{g}S^{1/2}}  \label{e3}
\end{equation}

\noindent is the energy contribution from the two dimensional Fermi gas\cite
{mca95}, with $g=2$ being the number of spin degrees of freedom for a spin $%
\frac 12$ fermion, and 
\begin{equation}
{\cal E}_C=2\pi \sigma ^2R^3=\frac{q^2}{8\pi R}=\frac{N^2e^2}{8\pi R}=\sqrt{%
\pi }\alpha \frac{N^2}{S^{1/2}}  \label{e4}
\end{equation}

\noindent is the Coulomb energy for the bubble with surface charge density $%
\sigma =q/(4\pi R^2)$ and $\alpha =\frac{e^2}{4\pi }$. (As in the Fermi ball
model, we also make the assumption that a fermion-antifermion asymmetry
exists so that fermions within a bubble wall do not all annihilate away.)
The bubble mass 
\begin{equation}
{\cal E}=\Sigma S+\left[ \frac{4\sqrt{\pi }N^{3/2}}{3\sqrt{g}}+\alpha \sqrt{%
\pi }N^2\right] S^{-1/2}  \label{e5}
\end{equation}

\noindent can be minimized with a surface area of 
\begin{equation}
S=4\pi R^2=\left\{ \frac 1{2\Sigma }\left[ \frac{4\sqrt{\pi }N^{3/2}}{3\sqrt{%
g}}+\alpha \sqrt{\pi }N^2\right] \right\} ^{2/3},  \label{e6}
\end{equation}

\noindent corresponding to an equilibrium radius of 
\begin{equation}
R=\left( \frac 1{4\pi }\right) ^{1/2}\left\{ \frac 1{2\Sigma }\left[ \frac{4%
\sqrt{\pi }N^{3/2}}{3\sqrt{g}}+\alpha \sqrt{\pi }N^2\right] \right\} ^{1/3}.
\label{e7}
\end{equation}

\noindent Note that the surface area independent ratio of the Coulomb energy
to Fermi gas energy is 
\begin{equation}
\frac{{\cal E}_C}{{\cal E}_F}=\frac 34\sqrt{g}\alpha N^{1/2}\approx \alpha
N^{1/2}.  \label{e8}
\end{equation}

From (\ref{e6}) we can write the surface area of the bubble as 
\begin{equation}
S=\frac 1{2\Sigma }\left( \frac{4\sqrt{\pi }N^{3/2}}{3\sqrt{g}S^{1/2}}+\frac{%
\alpha \sqrt{\pi }N^2}{S^{1/2}}\right) =\frac 1{2\Sigma }({\cal E}_F+{\cal E}%
_C).  \label{e9}
\end{equation}

\noindent The configuration energy is ${\cal E}=\Sigma S+{\cal E}_F+{\cal E}%
_C$, so that by (\ref{e5}) and (\ref{e9}), 
\begin{equation}
{\cal E}=\Sigma \left[ \frac 1{2\Sigma }({\cal E}_F+{\cal E}_C)\right] +%
{\cal E}_F+{\cal E}_C=\frac 32({\cal E}_F+{\cal E}_C).  \label{e10}
\end{equation}

\noindent [Also, by (\ref{e9}) ${\cal E}_F+{\cal E}_C=2\Sigma S=2{\cal E}_W$%
, so that ${\cal E=E}_W+{\cal E}_F+{\cal E}_C=3{\cal E}_W$.]

\subsection{Limiting Cases}

We can consider the limiting cases where either the Fermi gas contribution
dominates the Coulomb contribution, or vice versa, the Coulomb energy
dominates the Fermi gas energy. Let these cases be referred to as ``Fermi
gas dominance'' and ``Coulomb dominance'', respectively. For Fermi gas
dominance $\frac{{\cal E}_C}{{\cal E}_F}\ll 1$, which by (\ref{e8}) implies
that $\alpha N^{1/2}\ll 1$, whereas for Coulomb dominance $\frac{{\cal E}_C}{%
{\cal E}_F}\gg 1$, which implies that $\alpha N^{1/2}\gg 1$. Therefore a
stable bubble with a sufficiently {\it small} number of fermions will be
Fermi gas dominated with a mass ${\cal E}\approx \frac 32{\cal E}_F$, while
one with a sufficiently {\it large} number of fermions will be Coulomb
dominated with ${\cal E}\approx \frac 32{\cal E}_C$.

In order for the thin wall approximation to be respected, we require that $%
R/w\gg 1$, where $w=1/(\lambda \eta )$, and we assume for simplicity that $%
\lambda $ is of order unity. Therefore, in the thin wall approximation, the
equilibrium radius of the bubble must satisfy $R\eta \gg 1$. Let us take $%
R\eta \gtrsim \alpha ^{-1}$, so that from (\ref{e7}) we have (dropping
factors of order unity), 
\begin{equation}
R\eta \sim \left( \frac 1{4\pi }\right) ^{1/2}N^{1/2}\left[ 1+\alpha
N^{1/2}\right] ^{1/3}\gtrsim \alpha ^{-1},  \label{e11}
\end{equation}

\noindent from which we conclude that, roughly, $\alpha N^{1/2}\gtrsim O(1)$%
. Therefore, Fermi gas dominance is {\it not} realized for a stable,
spherical, thin walled bubble with a single species of electrically charged
fermion trapped within the wall, and Coulombic effects are therefore
necessarily nonnegligible. However, in what follows we can consider two
limits: (1) $\alpha N^{1/2}\sim 1$, in which case the Fermi gas and Coulomb
energy contributions are of comparable magnitude, and (2) $\alpha N^{1/2}\gg
1$, in which case the bubble is Coulomb dominated.

Although we have assumed that a mechanical equilibrium has been established
for the bubble, we must check for electrodynamic stability against charge
evaporation and also for stability against fermion emission from the wall,
i.e. the attractive force pulling a fermion into the wall must be larger
than the forces that would otherwise squeeze the fermion out of the wall
(e.g. the tension in the domain wall, the Coulomb force, and the Fermi gas
pressure) allowing the bubble to contract to a smaller radius.

\subsection{Stability Against Fermion Emission}

Consider a static bubble with fermion number $N\gg 1$ and mass ${\cal E}%
^{(N)}$. For it to be stable against releasing a fermion with mass $m_F$ in
the vacuum, we require that ${\cal E}^{(N+1)}-{\cal E}^{(N)}=\delta {\cal E}%
<m_F$. In the case of Coulomb dominance, the $N$-dependent energy
contributions in ${\cal E}$ are power functions of $N$, and we have,
approximately, for $N\gg 1$, $\delta {\cal E}\sim \frac{\partial {\cal E}}{%
\partial N}\delta N$. Therefore, for the Coulomb dominated bubble, from (\ref
{e9}) and (\ref{e10}) we have ${\cal E}\sim {\cal E}_C\sim \alpha
N^2/S^{1/2}\sim \alpha N^2(\Sigma ^{1/2}/{\cal E}_C^{1/2})$ which implies
that ${\cal E}_C\sim \alpha ^{2/3}N^{4/3}\Sigma ^{1/3}$, so that $\partial 
{\cal E}_C/\partial N\sim {\cal E}_C/N$. Setting $\delta N=1$ we then have
that $\delta {\cal E}<m_F$ implies that, roughly, 
\begin{equation}
{\cal E}<Nm_F  \label{e12}
\end{equation}

\noindent for stability against fermion emission. The fermion mass (in
vacuum) is $m_F=G_F\eta $, the domain wall surface energy is $\Sigma \sim
\eta ^3$, and for the Coulomb dominated bubble ${\cal E}\sim \alpha
^{2/3}N^{4/3}\eta $, so that (\ref{e12}) implies that 
\begin{equation}
(\alpha ^2N)^{1/3}<G_F=\left( \frac{m_F}\eta \right) .  \label{e13}
\end{equation}

Therefore, for the Coulomb dominance condition $\alpha ^2N\gg 1$ and (\ref
{e13}) to be simultaneously satisfied, we require 
\begin{equation}
1\ll \alpha ^2N<G_F^3.  \label{e14}
\end{equation}

\noindent The condition given by (\ref{e14}) can be met for a sufficiently
large fermion-scalar coupling $G_F$, but for a weaker coupling with $G_F$ of
order unity or smaller, a Coulomb dominated bubble apparently will not
stabilize when the wall is inhabited by a single species of charged fermion.

For the Fermi gas-Coulomb balance case $\alpha ^2N\sim 1$, we have ${\cal E}%
_C/{\cal E}_F\sim 1$ and by (\ref{e6}) $S\sim (\alpha \eta )^{-2}$, so that
by (\ref{e4}) and (\ref{e10}) ${\cal E}=\frac 32({\cal E}_F+{\cal E}_C)\sim 3%
{\cal E}_C\sim \alpha ^{-2}\eta $. Then, since $\alpha ^{-2}\sim N$, we get $%
{\cal E}^{(N)}\sim N\eta $ and $\delta {\cal E}\sim \eta $. In this case the
bubble is stable against fermion emission if 
\begin{equation}
\eta <m_F\Longrightarrow G_F>1.  \label{e14a}
\end{equation}

\noindent Thus, the requirement imposed upon $G_F$ by stability against
fermion emission seems to be a little more relaxed for the Fermi gas-Coulomb
balance case $\alpha N^{1/2}\gtrsim 1$ than for the Coulomb dominance case $%
\alpha N^{1/2}\gg 1$.

\subsection{Stability Against Charge Evaporation}

Although the bubble may be stable against fermion emission, it may not be
stable against charge evaporation. Consider, for example, a region where the
electric field strength is large enough to separate a virtual
electron-positron pair by a distance $r\sim 2m^{-1}$, where $m$ is the
electron mass, and bring the particles onto the mass shell. The work done by
the external electric field $eEr$ will be $\gtrsim 2m$ for a field strength $%
E\gtrsim m^2/e$, allowing the field to create $e^{\pm }$ pairs from the
vacuum. Therefore, if the electric field just outside the charged bubble is
above a critical value of $E_c\sim m^2/e$, where $m$ is the mass of a
charged particle, then the pair creation of charged particles becomes
probable. For a subcritical field strength $E<E_c$, pair creation from the
vacuum is suppressed. Taking $m$ to be the electron mass, we see that if the
electric field at the bubble surface becomes supercritical, i.e. $E>E_c$,
then electron-positron pairs are created from the vacuum, and a positively
charged bubble attracts electrons and repels positrons. The electrons
partially neutralize the bubble charge, thereby reducing the field until it
reaches a subcritical value $E<E_c$, so that the initial bubble charge
effectively evaporates through positron emission.

The electric field at the bubble's surface is $E=\sigma =Ne/S$, which is to
be compared to the critical field strength $E_c\sim m^2/e$. Now, for the
case of a Coulomb dominated bubble ($\alpha N^{1/2}\gg 1$) at equilibrium,
the surface area, from (\ref{e6}), is $S\sim \alpha ^{2/3}N^{4/3}/\Sigma
^{2/3}\sim \alpha ^{2/3}N^{4/3}/\eta ^2$, so that 
\begin{equation}
\frac E{E_c}\sim \frac{Ne/S}{m^2/e}=\frac{\alpha N}{4\pi m^2S}\sim \frac 
\alpha {4\pi }\left( \frac 1{\alpha ^2N}\right) ^{1/3}\left( \frac \eta m%
\right) ^2.  \label{e15}
\end{equation}

\noindent The field of a Coulomb dominated bubble will be subcritical for $%
(\alpha ^2N)^{1/3}>\frac \alpha {4\pi }\left( \frac \eta m\right) ^2$, so
that, in the case of Coulomb dominance, 
\begin{equation}
E<E_c\Longrightarrow \left( \frac \eta m\right) ^2<\left( \frac{4\pi }\alpha
\right) (\alpha ^2N)^{1/3}\text{ }.  \label{e16}
\end{equation}

\noindent Combining this condition with that of (\ref{e14}) then implies
that 
\begin{equation}
\left( \frac \eta m\right) ^2<\left( \frac{4\pi }\alpha \right) (\alpha
^2N)^{1/3}<\left( \frac{4\pi }\alpha \right) G_F\text{ },  \label{e16con}
\end{equation}

\noindent which, for a given value of $G_F$, places an upper limit on the
symmetry breaking energy scale $\eta $.

On the other hand, for the Fermi gas-Coulomb balance case $\alpha ^2N\sim 1$%
, 
\begin{equation}
\frac E{E_c}\sim \frac{e\eta ^2}{m^2/e}=4\pi \alpha \left( \frac \eta m%
\right) ^2,  \label{e16a}
\end{equation}

\noindent and for the field to be subcritical, 
\begin{equation}
E<E_c\Longrightarrow \left( \frac \eta m\right) ^2<\frac 1{4\pi \alpha }.
\label{e16b}
\end{equation}

\noindent In either case, unless the $Z_2$ symmetry breaking energy scale is
sufficiently close to the electron mass so that either (\ref{e16con}) or (%
\ref{e16b}) can be satisfied, a bubble containing only one species of
positively (negatively) charged fermion will undergo charge evaporation by
positron (electron) emission, reducing the net bubble charge to a value that
renders the external electric field subcritical. During such a process the
bubble will also change its lepton number. The stable bubble will therefore
be inhabited by {\it two} species of charged fermions, each contributing a
Fermi gas energy term, but the Coulomb energy term will be lowered.

In summary, a charged, thin walled, stable vacuum bubble populated with a
single charged fermion species will not be Fermi gas dominated, but a thin
walled bubble may stabilize if it is either Coulomb dominated or if there is
a Fermi gas - Coulomb balance. However, stability against fermion emission
and charge evaporation requires that (i) there be a sufficiently large
fermion coupling $G_F$ and (ii) that the symmetry breaking energy scale $%
\eta $ be sufficiently small. This last condition may be easily violated for
a value of $\eta $ on the order of the electroweak scale or higher, in which
case we expect stable charged vacuum bubbles to be populated with at least
two species of fermions, if stable bubbles exist at all.

\section{Vacuum Bubbles with Two Fermion Species - Neutral and Near-Critical
Charged Bubbles}

When the conditions for a subcritical electric field at a bubble's surface
can not be reached at equilibrium, so that electric field at the surface of
a charged bubble reaches a critical value $\sigma _{crit}\sim m^2/e$, where $%
m$ is the electron mass, then electron-positron pairs are produced resulting
in charge evaporation through electron absorption and positron emission,
until the electric field at the bubble surface drops slightly below its
critical value. (For definiteness, we take the charge of the heavy fermion
attached to the domain wall to be $+e$.) Since the pair production is
strongly suppressed for $\sigma <\sigma _{crit}$, we expect the bubble to
equilibrate by adjusting its radius, keeping its surface charge density
slightly subcritical. Therefore, for a bubble that has not been completely
neutralized, we take the surface charge density to be approximately constant
during the equilibration process, $\sigma \sim \sigma _{crit}\sim m^2/e$. Of
course, if the bubble is completely neutralized, $\sigma =0$, in which case
there are as many electrons in the bubble wall as there are heavy positively
charged fermions, i.e. $N_e=N_F$, where $N_e$ is the number of electrons in
the bubble and $N_F$ is the number of heavy fermions. For $N_e<N_F$, we have 
$\sigma =%
{\displaystyle {(N_F-N_e)e \over 4\pi R^2}}
\sim \sigma _{crit}$ which implies that the electron number 
\begin{equation}
N_e\sim N_F-\frac{4\pi \sigma _{crit}}eR^2\sim N_F-\frac{m^2}\alpha R^2
\label{e17}
\end{equation}

\noindent varies with the bubble radius $R$.

For a bubble that is microscopic in size, we expect the electron Fermi gas
to be relativistic. For this type of bubble containing two species of
charged fermion, the Fermi gas energy increases and the Coulomb energy
decreases as $N_e$ increases for a given value of $R$. In the Fermi gas
energy term of (\ref{e3}) we have $N^{3/2}\rightarrow N_F^{3/2}+N_e^{3/2}$,
while in the Coulomb energy term of (\ref{e4}) we have $N\rightarrow N_F-N_e$%
.

\subsection{Neutral Bubbles}

For the case that the stabilized bubble has been completely neutralized,
i.e. $N_e=N_F$, the Coulomb energy term vanishes, so that the configuration
energy of a spherical bubble is 
\begin{equation}
{\cal E}={\cal E}_W+{\cal E}_F=4\pi \Sigma R^2+\frac{4N_F^{3/2}}{3\sqrt{g}R}
\label{e18}
\end{equation}

\noindent giving an equilibrium radius of 
\begin{equation}
R=\left( \frac 1{6\pi \sqrt{g}\Sigma }\right) ^{1/3}N_F^{1/2}\sim \left( 
\frac 1{6\pi \sqrt{g}}\right) ^{1/3}\frac{N_F^{1/2}}\eta  \label{e19}
\end{equation}

\noindent and a bubble mass 
\begin{equation}
{\cal E}\sim 4\pi N_F\eta  \label{e20}
\end{equation}

\noindent For a thin-walled bubble ($\eta R\gg 1$) Eq. (\ref{e19}) implies
that $N_F\gg 1$.

As in the case of a single fermion bubble, we can examine the conditions
under which the neutral bubble will be stable against emission of heavy
fermions. We again require (assuming that $N_F\gg 1$) that ${\cal E}^{(N+1)}-%
{\cal E}^{(N)}=\delta {\cal E}\sim \partial {\cal E}/\partial
N_F<m_F=G_F\eta $. For the neutral bubble, ${\cal E}\sim 4\pi N_F\eta $, so
that for the bubble to be stable against heavy fermion emission we must have 
$G_F\gtrsim 4\pi $. For a fermion coupling much smaller than this, it
becomes energetically favorable for the heavy fermions to be expelled from
the bubble as it collapses, thus preventing stabilization.

However, although the neutral bubble can stabilize with a finite surface
area, as in the case with uncharged false vacuum bags the bubble is not
stable against flattening, so that as in the Fermi ball scenario, the bubble
can ultimately fragment into many small Fermi balls, each with a radius of
roughly $R_0\sim \eta ^{-1}$ at which point the fragmentation process stops.
(If there were a false vacuum volume energy term ${\cal E}_V\sim \Lambda V$
due to a slight breaking of the vacuum degeneracy, as is the case in the
original Fermi ball model, the tendency to fragment would be enhanced.)
These massive, neutral Fermi balls could serve as candidates for cold dark
matter.

Finally, let us note that a Fermi ball sized bubble with radius $R_0\sim
\eta ^{-1}$ can not be electrically charged if (i) $\eta /m\gg 1$ and (ii) $%
\alpha (N_F-N_e)\sim O(1)$ (as in the original Fermi ball model), since the
surface electric field would be supercritical in that case. This can be seen
by writing $R^2=(N_F-N_e)e/(4\pi \sigma )$ and noting that for $\sigma
\lesssim \sigma _{crit}\sim m^2/e$, the minimum radius the bubble with a
subcritical electric field can have is $R_{\min }\sim [(N_F-N_e)\alpha
]^{1/2}/m$. Thus, $R_{\min }/R_0\sim [(N_F-N_e)\alpha ]^{1/2}(\eta /m)$ is
not near unity and, consequently, we have that $R_{\min }\gg R_0$ if we
allow $\eta /m\gg 1$ while keeping $\alpha (N_F-N_e)$ roughly to an order of
unity. On the other hand, for $\alpha (N_F-N_e)\sim (m/\eta )^2$, we could
have $R_{\min }\sim R_0$, but such a bubble would have a charge number $%
(N_F-N_e)\sim \frac 1\alpha (m/\eta )^2<1$ for $(\eta /m)\gtrsim 1/\sqrt{%
\alpha }$, i.e., the bubble would have to be effectively neutral for $(\eta
/m)\gg 1$.

\subsection{Near-Critical Charged Bubbles}

For the case $N_F-N_e>0$, let us assume that the surface charge density is
near-critical, i.e. $\sigma \approx \sigma _{crit}\sim m^2/e$. Since $%
0<N_e<N_F$, we take the Fermi gas energy to be 
\begin{equation}
{\cal E}_F=\frac 2{3\sqrt{g}R}(N_F^{3/2}+N_e^{3/2})\sim \frac{N_F^{3/2}}R
\label{e21}
\end{equation}

\noindent The Coulomb energy is 
\begin{equation}
{\cal E}=2\pi \sigma ^2R^3\sim \frac{m^4R^3}\alpha ,  \label{e22}
\end{equation}

\noindent where $\alpha =e^2/4\pi $ and $m$ is the electron mass. For a
spherical bubble stabilized at a radius $R$, we have 
\begin{equation}
\frac{{\cal E}_C}{{\cal E}_F}\sim \frac{(mR)^4}{\alpha N_F^{3/2}},
\label{e23}
\end{equation}

\noindent which can be rewritten as 
\begin{equation}
R^2\sim \frac{\sqrt{\alpha }N_F^{3/4}}{m^2}\left( \frac{{\cal E}_C}{{\cal E}%
_F}\right) ^{1/2}.  \label{e24}
\end{equation}

Several limiting cases can be considered.

(1) {\it Fermi Gas Dominance}

\begin{equation}
\frac{{\cal E}_C}{{\cal E}_F}\ll 1\Rightarrow R^2\ll \frac{\sqrt{\alpha }%
N_F^{3/4}}{m^2}  \label{e25}
\end{equation}

(2) {\it Coulomb Dominance}

\begin{equation}
\frac{{\cal E}_C}{{\cal E}_F}\gg 1\Rightarrow R^2\gg \frac{\sqrt{\alpha }%
N_F^{3/4}}{m^2}  \label{e26}
\end{equation}

(3) {\it Fermi Gas-Coulomb Balance}

\begin{equation}
\frac{{\cal E}_C}{{\cal E}_F}\sim 1\Rightarrow R^2\sim \frac{\sqrt{\alpha }%
N_F^{3/4}}{m^2}  \label{e27}
\end{equation}

We write the total configuration energy of the bubble {\it at equilibrium}
(dropping factors of order unity) as 
\begin{equation}
{\cal E}={\cal E}_W+{\cal E}_F+{\cal E}_C\sim 4\pi \Sigma R^2+\frac{N_F^{3/2}%
}R+\frac{m^4R^3}\alpha  \label{e28}
\end{equation}

\noindent The equilibrium radius of the bubble is determined by a balance of
the radial forces acting on the bubble wall. Caution must be taken here to
not simply minimize ${\cal E}$ in (\ref{e28}), which would give a radially
inward Coulombic force, but rather to determine the force by considering a 
{\it virtual} displacement of the bubble wall, holding the charge $Q$ on the
wall fixed. (The charge of the bubble will vary with $R$, and hence time, in
general, as the bubble changes its radius during the physical equilibration
process. We can view $Q\approx 4\pi R^2\sigma _{crit}$ as a constraint on
the charge $Q$. We find the radial force at an instant by holding $Q$ fixed,
and considering how the energy of the configuration varies with $R$ at this
instant. The final result for the force at this instant is then obtained by
using the constraint $Q\approx 4\pi R^2\sigma _{crit}$.) For a {\it virtual}
change in the bubble's radius, holding the bubble charge $Q$ fixed at any
instant of time, we have ${\cal E}_C=Q^2/(8\pi R)$, and a virtual change in
energy due to a change in the radius alone is $\delta {\cal E}_C=-Q^2/(8\pi
R^2)\delta R$, allowing us to identify the radial electrostatic force by $%
F_R=-\delta {\cal E}/\delta R=Q^2/(8\pi R^2)$. Inserting the charge $%
Q\approx 4\pi R^2\sigma _{crit}$ into the expression for $F_R$ gives 
\begin{equation}
F_R\sim \frac{m^4R^2}{2\alpha },  \label{e29}
\end{equation}

\noindent which is a radially outward force tending to stabilize the bubble
against contraction. The total radial force on the bubble at equilibrium is
given by 
\begin{equation}
-F_R^{Tot}\sim 8\pi \Sigma R-\frac{N_F^{3/2}}{R^2}-\frac{m^4R^2}{2\alpha }%
\approx 0.  \label{e30}
\end{equation}

\noindent Taking $\Sigma \sim \eta ^3$, we have 
\begin{equation}
8\pi \eta ^3R^3-N_F^{3/2}-\frac{m^4R^4}{2\alpha }\approx 0.  \label{e31}
\end{equation}

\noindent Each of the limiting cases can be examined separately.

\smallskip\ 

(1) {\it Fermi Gas Dominance}

In this case the Fermi gas contribution to the force and configuration
energy is assumed to be much larger in magnitude than the Coulomb
contribution. The equilibrium radius is determined by $8\pi \eta ^3R^3\sim
N_F^{3/2}$, giving an equilibrium bubble radius 
\begin{equation}
R\sim \frac{N_F^{1/2}}\eta .  \label{e32}
\end{equation}

\noindent This is compatible with the condition given by (\ref{e25})
provided that $N_F^{1/4}\ll \eta ^2/m^2$. For a thin walled bubble ($\eta
R\gg 1$), we therefore require 
\begin{equation}
1\ll N_F\ll \left( \frac \eta m\right) ^8.  \label{e33}
\end{equation}

\noindent From (\ref{e17}) we find the electron number for the bubble to be $%
N_e\sim N_F\left[ 1-\alpha (m^2/\eta ^2)\right] $, which for $\alpha (m/\eta
)\ll 1$ gives $N_e\sim N_F$. The bubble mass is given by ${\cal E}/\eta \sim
N_F$. For the bubble to be stable against heavy fermion emission, we require
that $G_F\gtrsim 1$.

\smallskip\ 

(2) {\it Coulomb Dominance}

The bubble radius obtained from $8\pi \eta ^3R^3-m^4R^4/(2\alpha )\approx 0$
is 
\begin{equation}
R\sim \frac{\eta ^3}{m^4}.  \label{e34}
\end{equation}

\noindent This is compatible with (\ref{e26}) if $N_F^{1/4}\ll \eta ^2/m^2$,
and the bubble respects the thin wall approximation if $\eta R\sim (\eta
/m)^4\gg 1$. The bubble mass ${\cal E}\sim {\cal E}_W+{\cal E}_C$ is roughly
given by ${\cal E}/\eta \sim (\eta /m)^8$. This mass expression is
independent of $N_F$, since the Fermi gas term has been neglected, but from (%
\ref{e28}) we have that $\delta {\cal E}\sim \partial {\cal E}/\partial
N_F\sim N_F^{1/2}/R\sim (m^4/\eta ^3)N_F^{1/2}$, so that $\delta {\cal E}%
/\eta <G_F$, i.e. for the bubble to be stable against heavy fermion
emission, we have $G_F\gtrsim N_F^{1/2}(m/\eta )^4$. Since $N_F^{1/2}\ll
(\eta /m)^4$, the constraint on $G_F$ can be satisfied for $G_F\gtrsim 1$.

\smallskip\ 

(3) {\it Fermi Gas-Coulomb Balance}

We consider the Fermi gas force to be comparable to the Coulomb force in
this case, and therefore require $8\pi \eta ^3R^3\sim N_F^{3/2}$ and $%
N_F^{3/2}\sim m^4R^4/(2\alpha )$. The bubble radius is roughly 
\begin{equation}
R\sim \frac{\eta ^3}{m^4},  \label{e35}
\end{equation}

\noindent and $N_F\sim (\eta /m)^8$. The bubble mass is roughly given by $%
{\cal E}/\eta \sim N_F\sim (\eta /m)^8$. For stability against heavy fermion
emission, $G_F\gtrsim 1$.

\subsection{Black Hole Formation}

For a vacuum bubble to stabilize before forming a black hole, we require
that the stabilization radius $R$ be larger than the radius $R_H$ of the
outer horizon of the corresponding black hole state. For a neutral,
nonrotating black hole $R_H=2G{\cal E}$, where ${\cal E}$ is the black hole
mass, and for an extreme Reissner-Nordstrom black hole (with charge $Q=\sqrt{%
G}{\cal E}$) the outer horizon is located by $R_H=G{\cal E}$, so that for
the case of nonextemal or extremal nonrotating black holes we have, roughly, 
$R_H\sim G{\cal E}=\frac{{\cal E}}{M_P^2}$, where $M_P=(G)^{-1/2}$ is the
Planck mass. For the bubble to stabilize with a radius $R$ and avoid the
formation of a gravitationally collapsed black hole state we therefore
require that the equilibrium radius $R$ be larger than $R_H\sim \frac{{\cal E%
}}{M_P^2}$.

For the case of a Fermi gas dominated bubble, 
\begin{equation}
\frac R{R_H}\sim \left( \frac{M_P}\eta \right) ^2\frac 1{N_F^{1/2}},
\label{e36}
\end{equation}

\noindent so that for this bubble to avoid black hole formation, the number
of heavy fermions must be smaller than $N_{F,\max }\sim (M_P/\eta )^4$. The
maximum size and mass of such a bubble would then be, respectively, 
\begin{equation}
R_{\max }\sim \frac{N_{F,\max }^{1/2}}\eta \sim \left( \frac{M_P}\eta
\right) ^2\frac 1\eta ,  \label{e37}
\end{equation}
\begin{equation}
{\cal E}_{\max }\sim N_{F,\max }\eta \sim \left( \frac{M_P}\eta \right)
^4\eta .  \label{e38}
\end{equation}

\noindent (For a bubble that stabilizes at a GUT scale value $\eta \sim
10^{16}$GeV, for example, we have a maximum bubble radius $R_{\max }\sim
10^{-10}$GeV$^{-1}$ and a maximum mass ${\cal E}_{\max }\sim 10^{28}$GeV$%
\sim 10$kg. On the other hand, for $\eta \sim 10^4$GeV, for example, $%
R_{\max }\sim 10^{26}$GeV$^{-1}\sim 10^{10}$m and ${\cal E}_{\max }\sim
10^{64}$GeV, describing a very massive compact astrophysical object with a
mass of roughly 10 million solar masses and a radius of roughly 100 solar
radii!) Bubbles larger than that allowed by (\ref{e37}) would evidently form
black hole states, since the stabilization radius would lie inside the
horizon. Similar results hold for the neutral vacuum bubble. Also notice
that for the Fermi gas dominated (charged) bubble, by (\ref{e33}) we have $%
N_{F,\max }\ll (\eta /m)^8$, which implies that 
\begin{equation}
\eta >(M_Pm^2)^{1/3}\sim 10^4\text{GeV.}  \label{e38a}
\end{equation}

\noindent Therefore, a Fermi gas dominated charged bubble can evidently
reach a stable equilibrium only for a value of the symmetry breaking scale
in excess of roughly $10^4$GeV. We conclude that such a bubble formed at the
electroweak scale $\eta \sim 10^2$GeV would necessarily collapse to a black
hole.

Here, it is interesting to note that for a sufficiently small value of $\eta 
$ (e.g. $\eta \sim 10^4$GeV), the Fermi gas dominated bubbles described
above can have sizes and masses that become comparable to those of {\it %
neutrino balls} \cite{Holdom1}, which are particular examples of {\it cosmic
balloons} \cite{Holdom2}. (The spherical domain wall of a cosmic balloon
entraps fermions within its volume that become heavy outside the balloon.)
However, a fundamental difference between a bubble and a neutrino ball (NB)
is that the domain wall of the neutrino ball is essentially transparent to
all matter and radiation, except for neutrinos, whereas the domain wall of
the bubble is not transparent, as it is inhabited by charged fermions.
Stars, gravitationally attracted to the neutrino ball, can simply drift
through the NB domain wall \cite{HM}. Inside the NB, there will be a
frictional force exerted on a star by the ambient neutrinos, so that
eventually the star will tend to reside at or near the center of the NB. The
ambient neutrino gas can speed the star's evolution, enhancing its
probability rate to undergo a supernova type of explosion. Holdom and
Malaney \cite{HM} have proposed a mechanism wherein the neutrino emissions
from such explosions within NBs can be converted into intense gamma ray
bursts. This mechanism, however, would not apply to a vacuum bubble, since
the domain wall of the bubble is not invisible to matter and radiation, and
hence stars can not simply drift through the domain wall of the bubble.
Furthermore, we have considered the case where vacuum, rather than fermionic
gas, occupies the bubble's interior.

For the case of either a Coulomb dominated bubble or a Fermi gas-Coulomb
balanced bubble, we have $R\sim \left( \frac \eta m\right) ^3\frac 1m$ and $%
{\cal E}\sim \left( \frac \eta m\right) ^8\eta $, so that 
\begin{equation}
\frac R{R_H}\sim \left( \frac m\eta \right) ^4\left( \frac{M_P}\eta \right)
^2.  \label{e39}
\end{equation}

\noindent Therefore, for one of these bubbles to avoid black hole formation
we must have $R>R_H$ which implies that 
\begin{equation}
\eta \lesssim (M_Pm^2)^{1/3}\sim 10^4\text{GeV.}  \label{e40}
\end{equation}

\noindent For sufficiently small symmetry breaking scales (e.g. the
electroweak scale, $\eta \sim 10^2$GeV), black hole formation is avoided,
but for values of $\eta $ much greater than that of (\ref{e40}) (e.g. the
GUT scale, $\eta \sim 10^{16}$GeV), black hole formation evidently can not
be avoided. (For a bubble that stabilizes at an electroweak scale value $%
\eta \sim 10^2$GeV, for example, we have a bubble radius and mass of $R\sim
10^{18}$GeV$^{-1}\sim 10^2$ m and ${\cal E}\sim 10^{42}$GeV$\sim 10^{15}$
kg, respectively.) Therefore, whether or not a particular type of charged
bubble can eventually stabilize depends upon whether $\eta $ is above or
below a value of roughly 10 TeV.

\section{Summary and Conclusions}

If either a {\it biased,} exact discrete symmetry or an {\it approximate}
discrete symmetry is spontaneously broken, a network of bounded domain wall
surfaces giving rise to ``vacuum bubbles{\it ''} may result. The dynamical
evolution of a bubble will depend upon what other fields couple to the
scalar field forming the domain wall. If fermions couple to the scalar field
in such a way that it becomes favorable for the fermions to reside within
the bubble wall, the resulting degenerate Fermi gas can help to stabilize
the bubble against an unchecked collapse. A scenario of this type
incorporating electrically neutral fermions plays an essential role in the
Fermi ball model\cite{mca95}, for example, where the domain wall forms a
thin skin enclosing a false vacuum. For a sufficiently strong fermion
coupling to the scalar field, the fermions remain within the wall and allows
the resulting bag-like configuration to equilibrate with a finite nonzero
surface area. In turn, this vacuum bag can flatten and fragment, resulting
in the production of many smaller ``Fermi balls''. Thus, the cosmological
domain wall problem can be evaded through the formation of a bubble network,
and ultimately, Fermi balls. Here, attention has been focused upon an
extension of this type of scenario, where the fermions are assumed to have
an electric U(1) gauge charge, introducing nontrivial electromagnetic
effects that must be taken into consideration when examining the stability
of a {\it charged} vacuum bubble. It has been demonstrated that the
Coulombic effects {\it can not} be considered negligible in comparison to
the Fermi gas effects for the case of a thin walled bubble inhabited by a
single species of charged fermion.

The physical realization of stable, static, charged vacuum bubbles also
depends upon two particular stability issues: (i) stability of the bubble
against an emission of the fermion coupled to the scalar field, and (ii)
stability against charge evaporation, which can occur when the surface
electric field of the bubble becomes too large, or supercritical. The
stability against fermion emission can, in many cases, be satisfied if the
fermion-scalar coupling occupies a range, given roughly by $G_F\gtrsim 1-10$%
, which may be regarded as fairly natural. However, for stability against
charge evaporation, a relatively severe constraint is placed upon the
symmetry breaking energy scale $\eta $. More specifically, unless the $Z_2$
symmetry breaking energy scale is sufficiently close to the electron mass,
so that either (\ref{e16con}) or (\ref{e16b}) can be satisfied, a bubble
initially containing only one species of positively (negatively) charged
fermion will undergo charge evaporation by positron (electron) emission,
reducing the net bubble charge to a value that renders the external electric
field subcritical. This may easily be the case for a value of $\eta $ on the
order of the electroweak scale or higher. During such a process the bubble
will also change its lepton number. The stable bubble will then be inhabited
by {\it two} species of charged fermions, each contributing a Fermi gas
energy term, but the Coulomb energy term will be lowered.

Therefore, consideration has subsequently been given to the case of a bubble
populated by {\it two} species of fermions - the heavy fermions coupling
directly to the domain wall scalar field, and electrons that have been
absorbed by the bubble in order to partially or completely neutralize the
bubble, rendering the surface electric field subcritical. The charge
evaporation allows the Coulombic effects to be diminished, while the Fermi
gas effects are enhanced. These two-fermion species bubbles may stabilize
through Fermi gas dominance, through Coulomb dominance, or through a Fermi
gas-Coulomb balance, depending upon the value of $\eta $ and final
configuration parameters, such as the bubble mass and the numbers of
fermions populating the bubble (see sec. 3). Finally, constraints have been
estimated for stable bubbles that do not form black holes. These constraints
take the form of limits for either the number of heavy fermions that can
populate a particular type of bubble (for the cases of neutral or charged,
Fermi gas dominated bubbles), and/or for the symmetry breaking scale $\eta $
(for the cases of near-critically charged bubbles). For instance, if $\eta
<10^4$GeV, then Fermi gas dominated bubbles do not stabilize before forming
black holes, whereas if $\eta >10^4$GeV, then Coulomb dominated or Fermi
gas-Coulomb balanced bubbles necessarily collapse into black holes.

In summary, it has been argued that if there existed a biased, or an
approximate, discrete symmetry which was broken in the early universe, and
if heavy charged fermions coupled to the domain wall-forming scalar field,
then it is possible for stable, charged vacuum bubbles to be produced,
provided that certain parameters, such as the symmetry breaking energy
scale, the fermion-scalar coupling constant, and fermion numbers, occupy
appropriate ranges. If the parameters do not lie within such ranges, the
bubbles are expected to undergo an unchecked collapse. Stable bubbles may
indeed form, but it is not known what fraction of bubbles will actually
stabilize, since this presumably depends upon how the data of initial
conditions are distributed over the collection of evolving bubbles. At any
rate, it is possible that such bubble configurations, even if rare, could be
physically realized, and the physical existence of such vacuum bubbles could
have interesting consequences for particle physics and cosmology.

\smallskip\ 

Acknowledgement:

\smallskip\ 

I thank D. Bazeia for discussions related to this work.

\smallskip\ 


\end{document}